\documentclass[lettersize,journal]{IEEEtran}
\usepackage{amsmath,amsfonts}
\usepackage{algorithmic}
\usepackage{algorithm}
\usepackage{array}
\usepackage[caption=false,font=normalsize,labelfont=sf,textfont=sf]{subfig}
\usepackage{textcomp}
\usepackage{stfloats}
\usepackage{color}
\usepackage{verbatim}
\usepackage{graphicx}
\usepackage{cite}
\usepackage[hyphens]{url}
\usepackage{breakurl}
\usepackage{textcomp}
\usepackage{subfig}
\usepackage{booktabs}
\usepackage{multirow}
\usepackage{amssymb}
\usepackage{pifont}

\newcommand{\cmark}{\ding{51}}

\begin{document}

\title{Biometrics in the Era of COVID-19:\\ Challenges and Opportunities}

\author{Marta Gomez-Barrero, Pawel Drozdowski, Christian Rathgeb, Jose Patino, Massimiliano Todisco, \\Andreas Nautsch, Naser Damer, Jannier Priesnitz, Nicholas Evans, Christoph Busch
\thanks{M. Gomez-Barrero is with the Hochschule Ansbach, Germany. P. Drozdowski, C. Rathgeb, J. Priesnitz, and C. Busch are with the da/sec - biometrics and internet security research group, Hochschule Darmstadt, Germany. J. Patino, M. Todisco, A. Nautsch, and N. Evans are with EURECOM, France. N. Damer is with the Fraunhofer Institute for Computer Graphics Research IGD, Germany.

This research work has been funded by the German Federal Ministry of Education and Research and the Hessian Ministry of Higher Education, Research, Science and the Arts within their joint support of the National Research Center for Applied Cybersecurity ATHENE and the DFG-ANR RESPECT Project (406880674 / ANR-18-CE92-0024).}
\thanks{Manuscript received December 1, 2021; revised July, 2022.}}

\markboth{}%
{Gomez-Barrero \MakeLowercase{\textit{et al.}}: Biometrics in the Era of COVID-19: Challenges and Opportunities}


\maketitle

\begin{abstract}
Since early 2020, the COVID-19 pandemic has had a considerable impact on many aspects of daily life. A range of different measures have been implemented worldwide to reduce the rate of new infections and to manage the pressure on national health services. A primary strategy has been to reduce gatherings and the potential for transmission through the prioritisation of remote working and education. Enhanced hand hygiene and the use of facial masks have decreased the spread of pathogens when gatherings are unavoidable. These particular measures present challenges for reliable biometric recognition, e.g.\ for facial-, voice- and hand-based biometrics. At the same time, new challenges create new opportunities and research directions, e.g.\ renewed interest in non-constrained iris or periocular recognition, touch-less fingerprint- and vein-based authentication and the use of biometric characteristics for disease detection. This article presents an overview of the research carried out to address those challenges and emerging opportunities.
\end{abstract}


\section{Introduction}
\label{sec:introduction}

Since early 2020, the world has been grappling with the COVID-19 pandemic caused by the new SARS-CoV-2 coronavirus. At the time of writing, there have been more than 250 million confirmed infections while more than five million have succumbed to the virus or related complications~\cite{WHO-COVID}. The main vector of disease transmission is exposure to respiratory particles resulting from direct or close physical contact with infected individuals. Transmission can also occur from the transfer of viral particles from contaminated surfaces or objects to the eyes, nose or mouth~\cite{WHO-COVID}.

\begin{figure}[t]
\centering
\subfloat[Surgical mask~\protect \cite{Damer-Masks-2020}]{\includegraphics[height=0.150\textwidth]{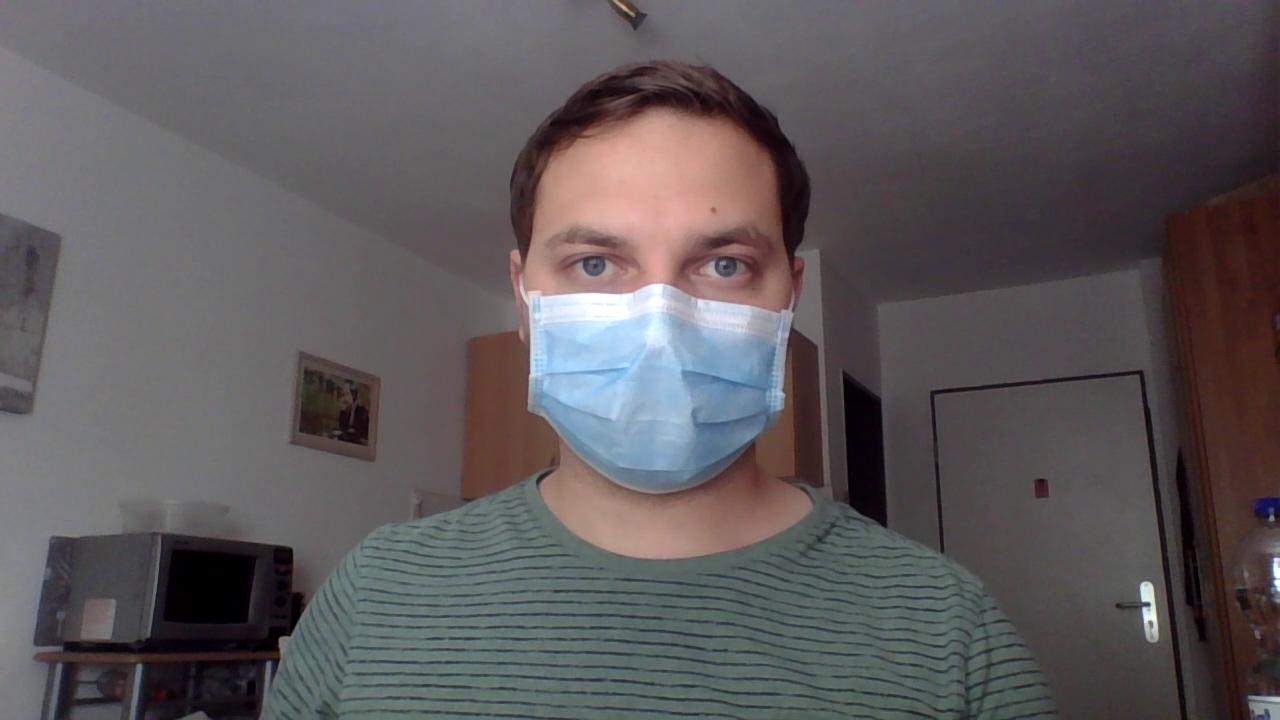}} \hfill
\subfloat[Cloth mask~\protect \cite{Damer-Masks-2020}]{\includegraphics[height=0.15\textwidth]{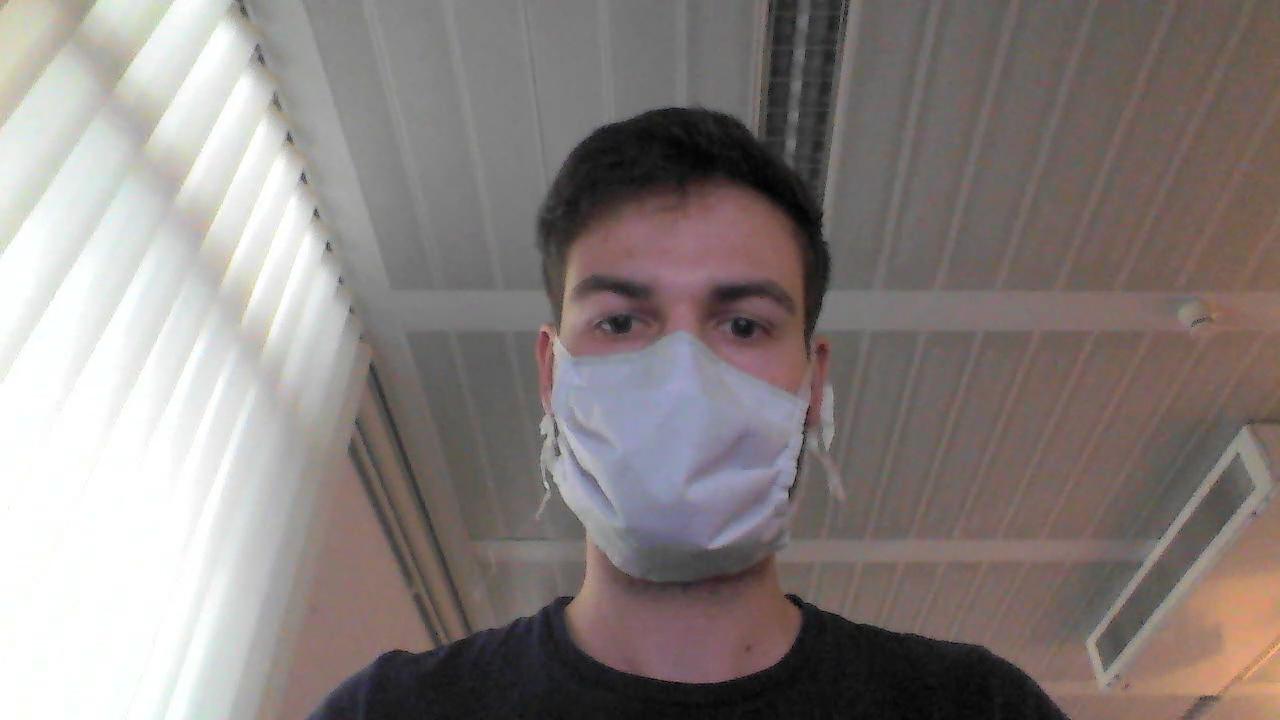}} \hfill
\subfloat[Filter mask\protect\footnotemark]{\includegraphics[height=0.15\textwidth]{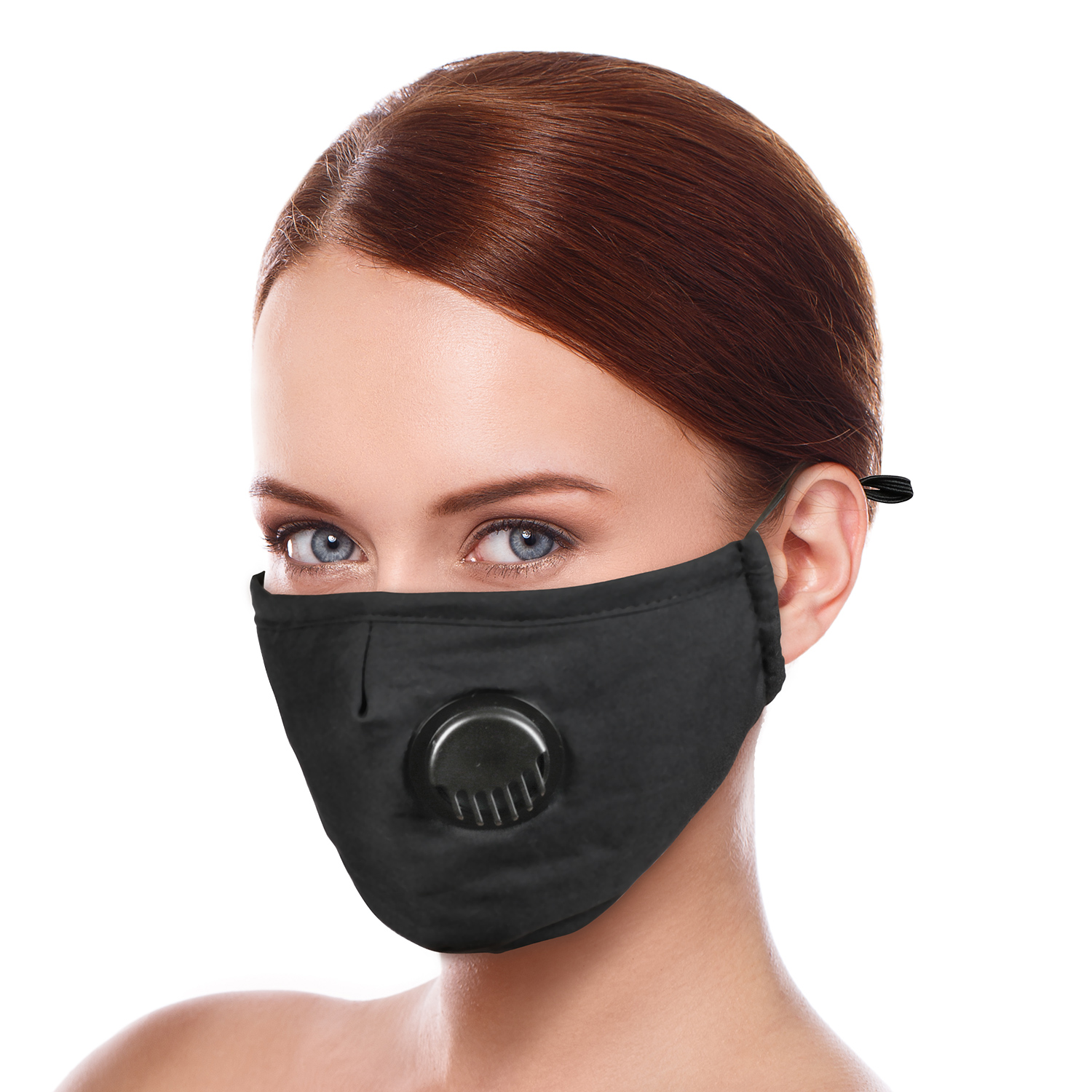}}
\subfloat[Printed mask\protect\footnotemark]{\includegraphics[height=0.15\textwidth]{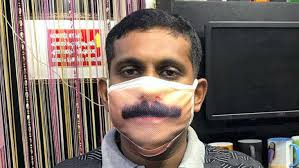}}
\caption{Examples of typical protective face masks}\vspace*{-0.5cm}
\label{fig:face_masks}
\end{figure}

Various preventive measures have been adopted worldwide to help curb the spread of the virus by reducing the risk of new infections. These include local, national and international travel restrictions, the banning of large gatherings and the encouragement of physical distancing, remote working and education, and strict quarantine policies, see e.g.~\cite{EU-Measures}. Two of the most broadly adopted measures are the (sometimes mandatory) use of protective facial coverings or masks~\cite{MasksCountries-2020} and enhanced hand hygiene (handwashing or disinfection using hydroalcoholic gel). Facial masks, such as those illustrated in Fig.~\ref{fig:face_masks}, can reduce viral transmission through respiratory particles~\cite{Peeples-FaceMasks-2020}, while enhanced hand hygiene can reduce the rate of new infections through contact with contaminated surfaces or objects. Preventive measures, as well as the virus itself, have necessitated consequential shifts and disruption to daily life, with potentially long-lasting repercussions impacting individuals, social and professional practices and processes, businesses both small and large, as well as the global economy.

Such measures have had a considerable impact in our daily lives. For instance, the use of facial masks covering the mouth and nose in public spaces can decrease the usefulness of surveillance systems or prevent us from unlocking our smartphone using face recognition technologies. In this context, this article focuses on the impact of the COVID-19 pandemic on \textbf{biometric recognition}. 
Biometric technologies can be used for automated identity verification and to distinguish individuals based on their personal biological and behavioural characteristics (e.g.\ face and voice). Biometric solutions frequently supplement or replace traditional knowledge- and token-based security systems since, as opposed to passwords and access cards, biometric characteristics cannot be forgotten or lost. Furthermore, biometrics inherently and seamlessly enable diverse application scenarios which are either difficult or infeasible using more traditional methods, e.g.\ continuous authentication~\cite{Patel-ContinuousAuthentication-2016,Mondal-ContinuousAuthentication-2017}, forensics~\cite{Tistarelli-HandbookBiometricForsensics-2017}, and surveillance~\cite{HandbookRemote}. 

\footnotetext[1]{Source: \url{www.ikatehouse.com}}
\footnotetext[2]{Source: \url{www.thenationalnews.com}}


\begin{table*}[t]
\centering
\caption{Overview of commonly used biometric characteristics in the context of COVID-19.}
\label{table:overview}
\resizebox{\textwidth}{!}{
\begin{tabular}{cccccccc}
\toprule
\textbf{Biometric} & \textbf{Data acquisition} & \multicolumn{4}{c}{\textbf{Application area}} & \textbf{Operational} & \textbf{Impact of} \\ \cmidrule{3-6}
\textbf{characteristic} & \textbf{hardware} & \textbf{mobile devices} & \textbf{access control} & \textbf{forensics} & \textbf{surveillance} & \textbf{prevalence} & \textbf{COVID-19} \\ \midrule 
Face & commodity hardware & \cmark & \cmark & \cmark & \cmark & wide & high\\ \midrule
NIR Iris & special sensor & (\cmark) & \cmark & & & wide & low\\
VIS Iris & commodity hardware & \cmark & (\cmark) & & & low & low\\ \midrule
Touch-based Fingerprint & special sensor & \cmark & \cmark & \cmark & & wide & high\\
Touchless Fingerprint  & commodity hardware & \cmark & \cmark & & & low & low\\ \midrule
Touch-based Hand Vein & special sensor & & \cmark & & & low & low\\
 Touchless Hand Vein & special sensor  & (\cmark) & \cmark & & & low & low \\ \midrule
Voice & commodity hardware & \cmark & \cmark & \cmark & \cmark & wide & medium\\
 \bottomrule
\end{tabular}
}
\end{table*}

Biometric technologies have come to play an integral role in society, e.g., for identity management, surveillance, access control, social and welfare management, and automatic border control, with these applications alone being used either directly or indirectly by billions of individuals~\cite{AlRaisi-UAEIris-Elsevier-2008,Dalwai-Aadhaar-2014,SmartBorders-EU-2018,Thales-AFIS-2020}. While reliance upon biometric technologies has reached a profound scale, health-related measures introduced in response to the COVID-19 pandemic have been shown to impact either directly or indirectly upon their reliability \cite{Carlaw-BTT-BiometricsCovid-2020}. It should be however noted that the new measures have a limited impact on other biometric characteristics such as ear \cite{Emersivic-UnconstrainedEarChallenge-ICB-2019}. Even though this fact will also lead to renewed efforts directed to such biometric characteristics in order to achieve accurate and deployable systems in the near future, we limit the scope of this article to those biometric characteristics affected by health-related measures.

Table~\ref{table:overview} provides a brief overview of the operational prevalence and COVID-19-related impacts and technological challenges in the context of the most widely (in operational systems) used biometric characteristics. They are reviewed and discussed in further detail in the remainder of this article, including a short introduction and description for each characteristic for the non-expert readers. This work represents a narrative/integrated review. It is meant to selectively assess relevant works in the field of biometrics that (in)directly tackle challenges caused by the COVID-19 pandemic. It is aiming at offering guidance about future research directions and enabling new perspectives to emerge.

The rest of the article is organised as follows. The impact of facial masks on biometrics technologies is discussed in Section~\ref{sec:influence}. Section~\ref{sec:remote_biometric_authentication} addresses impacts upon mobile and remote biometric authentication. Section~\ref{sec:emerging_technologies} describes new opportunities and applications that have emerged as a result of the COVID-19 pandemic. The societal impact of these changes is discussed in Section~\ref{sec:societal} and concluding remarks are presented in Section~\ref{sec:conclusions}.

\vspace{0.5cm}

\section{Influence of facial coverings on biometric recognition}
\label{sec:influence}
The use of facial coverings, such as masks, occlude a substantial part of the lower face. Such occlusions or obstructions change dramatically the operational conditions for numerous biometric recognition technologies. Such changes can make biometric recognition especially challenging. A review of the impacts of facial coverings is presented in this section, with a focus upon facial, periocular, iris, and voice biometrics.

\subsection{Face recognition}
\label{subsec:influence_face}

The natural variation among individuals yields a good inter-class separation and thus makes the use of facial characteristics for biometric recognition especially appealing. Traditional solutions rely upon handcrafted features based on texture, keypoints, and other descriptors for face recognition~\cite{Li-HandbookFace-2011}. More recently, the use of deep learning and massive training datasets has led to breakthrough advances. The best systems perform reliably even with highly unconstrained and low-quality data samples~\cite{Masi-DeepFaceSurvey-2018,Guo-DeepFaceSurvey-2019}. Relevant to the study presented here is a large body of research on occluded face detection~\cite{Opitz-OccludedFaceDetection-2016} and recognition~\cite{Zeng-OccludedFaceSurvey-2020}, though occlusion-invariant face recognition remains challenging~\cite{Song-OcclusionRobustFaceRecognition-2019}. Most work prior to the COVID-19 pandemic addresses occlusions from, e.g., sunglasses, partial captures, or shadows which typify unconstrained, `in-the-wild' scenarios. The use of facial masks therefore presents a new and significant challenge to face recognition systems, especially considering the stringent operating requirements for application scenarios in which face recognition technology is often used, e.g. automated border control. The requirement for extremely low error rates typically depend on the acquisition of unoccluded images of reasonable quality.


The most significant evaluation of the impact of masks upon face recognition solutions was conducted by the National Institute of Standards and Technology (NIST)~\cite{Ngan-NIST-PreCovid-2020,Ngan-NIST-PostCovid-2020}. The evaluation was performed using a large dataset of facial images with superimposed, digitally generated masks of varying size, shape, and colour. 
The evaluation tested the face recognition performance of algorithms submitted to the ongoing Face Recognition Vendor Test (FRVT) benchmark in terms of biometric verification performance (i.e., one-to-one comparisons). The false-negative error rates (i.e., false non-match rate) for algorithms submitted prior to the pandemic~\cite{Ngan-NIST-PreCovid-2020}, where observed to increase by an order of magnitude, even for the most reliable algorithms. Even some of the best-performing algorithms (as judged from evaluation with unmasked faces) failed almost completely, with false-negative error rates of up to 50\%. 

Of course, these results may not be entirely surprising given that systems designed prior to the pandemic are unlikely to have been optimised for masked face data. The study itself also had some limitations, e.g.\ instead of using genuine images collected from mask-wearing individuals, it used synthetically generated images where masks were superimposed using automatically derived facial landmarks. Despite the shortcomings, the study nonetheless highlights the general challenges to biometric face recognition from face coverings and masks. The general observations are that: 1) the degradation in verification reliability increases when the mask covers a larger proportion of the face including the nose; 2) reliability degrades more for mated biometric comparisons than for non-mated comparisons, i.e.\ masks increase the rate of false non-match rate more than the false match rate; 3) different mask shapes and colors lead to differences in the impact upon verification reliability, a finding which emphasises the need for evaluation using genuine masked face data; 4) in many cases, masked faces are not even detected. 

A follow-up study~\cite{Ngan-NIST-PostCovid-2020}, also conducted by NIST, evaluated systems that were updated with enhancements designed to improve reliability for masked faces. In addition to greater variability in mask designs, the study also considered both masked probe as well as masked reference face images. While reliability was observed to improve for masked faces, it remained substantially degraded compared to unmasked faces (approximately an order of magnitude lower). The degraded performance of masked faces was equivalent to that for unmasked faces and state-of-the-art systems from 2017. Increases in false-match rates were also observed when both reference, as well as probe faces are masked. Full details and results are available from the NIST FRVT Face Mask Effects website~\cite{FRVT-Masks-2020}.

Results from the related DHS Biometric Rally show similar trends~\cite{DHS-Rally-2020}. The DHS study was conducted in a setup simulating real operational conditions using systems submitted by commercial vendors. Significant difficulties in image acquisition as well as general degradation in biometric performance were observed for masked faces. Like the NIST study, the DHS study too found that, even with masked faces, today's systems perform as well as state-of-the-art systems from only a few years ago~\cite{DHS-Rally-2020} tested with unmasked face images.


These US-based studies are complemented by a number of academic studies.
Two datasets~\cite{Damer-Masks-2020,Wang-MaskedFaceDB-2020} of masked face images have been collected in Europe and China to support research efforts. While~\cite{Wang-MaskedFaceDB-2020} provides data, however, it does not provide a formal evaluation of the effect of masks on face recognition performance. Moreover, this study did not address a specific usecase scenario, e.g. collaborative face verification. 
Damer \textit{et al.}~\cite{Damer-Masks-2020,DBLP:journals/iet-bmt/DamerBSKK21,https://doi.org/10.1049/bme2.12077} released a database of real masked face images that were collected in three collaborative sessions. They include realistic variation in the capture environment, masks, and illumination. Evaluation results show similar trends exposed by the NIST study~\cite{Ngan-NIST-PreCovid-2020}: difficulties in face detection and greater impacts upon mated comparisons than non-mated comparisons.
While significantly smaller than the NIST dataset in the number of data subjects and images, the use of real instead of synthetically generated masked faces images increases confidence in results. 


From a technical perspective, face masks can be considered as a subset of general face occlusions, and thus previous works on this issue are relevant. A number of works have proposed to automatically detect, and synthetically in-paint, the occluded face areas. This aimed at generating realistic and occlusion-free face images, as well as enabling a more accurate face recognition. Most of the better performing face completion solutions are based on deep generative models \cite{DBLP:conf/cvpr/LiLY017, DBLP:conf/accv/ZhangZS018}.
A recent study by Mathai \textit{et al.}~\cite{DBLP:conf/icb/MathaiMA19} has shown that face completion can be beneficial for occluded face recognition accuracy, given that the occlusions are detected accurately. They have also pointed out that the completion of occlusions on the face boundaries did not have significant effect, which is not the case of face mask occlusions. Thus, these results indicate that face image completion solutions are possible candidates to enhance masked face recognition performance.

\begin{figure}[t]
\centering
\subfloat[Transparent mask\protect\footnotemark]{\includegraphics[width=0.3\textwidth]{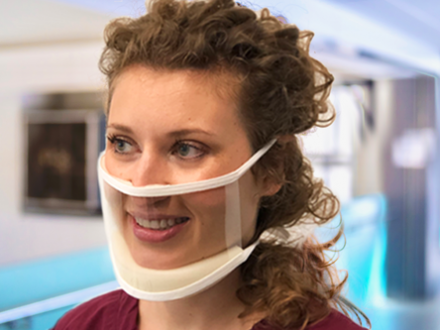}} \hfil
\subfloat[Face shield\protect\footnotemark]{\includegraphics[width=0.3\textwidth]{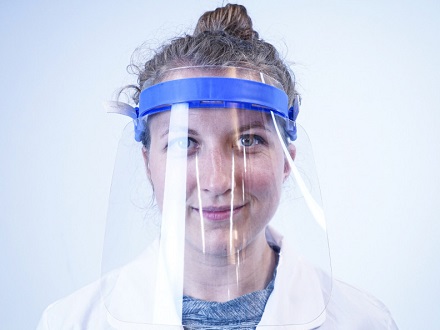}}
\caption{Examples of alternative protective masks}\vspace*{-0.5cm}
\label{fig:alternative_masks}
\end{figure}
\footnotetext[3]{Source: \url{https://www.theclearmask.com/product}}
\footnotetext[4]{Source: \url{https://3dk.berlin/en/covid-19/474-kit-for-face-shield-mask-with-two-transparent-sheets.html}}

The use of transparent masks or shields may combat to some extent the impact of opaque masks upon face recognition systems. Transparent masks, such as those shown in Fig.~\ref{fig:alternative_masks}, allow some portion of the masked face to remain visible but even their impact is likely non-trivial. Transparent masks can cause light reflections, visual distortions and/or blurring. Both opaque and transparent masks, as well as strategies to counter their impact, may increase the threat of presentation attacks. For example, it is conceivable that masks with specific patterns could be used to launch concealment or impersonation attacks, e.g.\ using concepts similar to those in~\cite{Sharif-AccessoriesImpersonationConcealment-2016}. 

Regardless of the exact type of face mask, wearing one can have an effect on the face image quality. Most biometric systems estimate the quality of a detected face image prior to feature extraction \cite{DBLP:journals/corr/abs-2009-01103}. This quality estimation indicates the suitability of the image for recognition purposes \cite{DBLP:conf/cvpr/TerhorstKDKK20,DBLP:journals/corr/abs-2112-06592}. For existing systems, the quality threshold configurations might lead to disregarding samples with face masks and thus increase the failure to extract rate. This link between face occlusions and face image quality has been probed in previous works, however, not exclusively for mask occlusions. One of these works, presented by Lin and Tang \cite{DBLP:conf/cvpr/LinT07}, built on the assumption that occlusions negatively effect the face image quality, in order to detect such occlusion. A recent study by Zhang \textit{et al.}~\cite{DBLP:conf/icct/ZhangSYDZS19} has demonstrated the effect of occlusion on the estimated face mage quality, along with presenting an efficient multi-branch face quality assessment algorithm. The authors pointed out that images with alignment distortion, occlusion, pose or blur tend to obtain lower quality scores.

The studies conducted thus far highlight the challenges to face recognition systems in the COVID-19 era and raise numerous open questions. These include, but are not limited to large-scale tests using images with real and not digitally generated masks, identification (i.e. one-to-many search), demographic differentials, presence of additional occlusions such as glasses, the effect on face image quality \cite{DBLP:conf/biosig/FuSCD21}, unconstrained data acquisition in general, as well as effects on the accuracy of human examiners~\cite{Ngan-NIST-PostCovid-2020,https://doi.org/10.1049/bme2.12077}. In addition, new areas of research have been opened, such as the automatic detection of whether a subject is wearing the mask correctly (i.e., covering mouth and nose) \cite{Batagelj-CorrectUseMasks-AS-2021}.

To foster research on the aforementioned issues, the Masked Face Recognition Competition (MFR) \cite{Boutros-MFRC-IJCB-2021} was organised in 2021. The main goals of this competition were not only the enhancement of recognition performance in the presence of masks, but also the analysis of the deployability of the proposed solutions. A private dataset representing a collaborative multi-session real masked capture scenario was used to evaluate the submitted solutions. In comparison to one of the top performing academic face recognition solutions, 10 out of the 18 submitted solutions did achieve a higher masked face verification accuracy, thereby showing the way for future face recognition approaches. This was followed by a series of works that targeted enhancing the accuracy of masked face recognition, either by training task-specific models \cite{DBLP:conf/fgr/HuberBKD21} or processing face templates extracted by existing models \cite{DBLP:journals/pr/BoutrosDKK22}.

\subsection{Iris recognition}
\label{subsec:influence_iris}
The human iris, an externally visible structure in the human eye, exhibits highly complex patterns which vary among individuals. The phenotypic distinctiveness of these patterns allow their use for biometric recognition~\cite{Daugman-HowIrisRecognitionWorks-IEEE-2004}. The acquisition of iris images typically requires a camera with near-infrared (NIR) illumination so that sufficient detail can be extracted for even darkly pigmented irides. Recent advances support acquisition in semi-controlled environments at a distance even from only reasonably cooperative data subjects on the move (e.g.\ while walking)~\cite{Matey2009,Nguyen-Iris-2017}.

Solutions to iris recognition which use mobile devices and which operate using only visible wavelength illumination have been proposed in recent years~\cite{Proenca-TPAMI-IrisVIS-2009,Raja-PRL-SmartphoneIrisVIS-2015,Rattani-IVC-OcularVISSurbvey-2017}. Attempts to use image super-resolution, a technique of generating high-resolution images from low resolution counterparts, have also shown some success by increasing image quality~\cite{Tapia-WIFS-SRIris-2020}. However, iris recognition solutions seem more dependent than face recognition solutions upon the use of constrained scenarios that lead to the acquisition of high quality images~\cite{Masi-DeepFaceSurvey-2018,Guo-DeepFaceSurvey-2019}. Nevertheless, iris recognition systems have now been in operation worldwide for around two decades. Near-infrared iris recognition has been adopted in huge deployments of biometrics technology, e.g.\ in the context of the Indian Aadhaar programme through which more than 1 billion citizens have been enrolled using iris images~\cite{UIDAI-Dashboard} in addition to other biometric data. Due to their high computational efficiency and reliability~\cite{Daugman-Doppelgangers-2016}, iris recognition systems are used successful within the Aadhaar programme for intensive identification (1-$N$ search) and de-duplication ($N$-$N$ search)~\cite{Dalwai-Aadhaar-2014}.

\begin{figure}[t]
\centering
\includegraphics[width=0.75\linewidth]{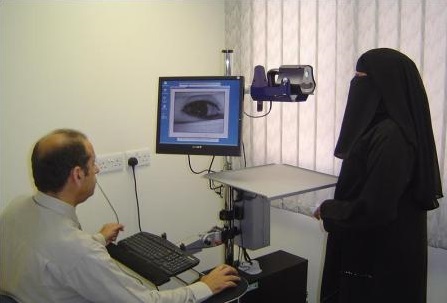}
\caption{IrisGuard Inc. UAE enrolment station\protect\footnotemark}\vspace*{-0.5cm}
\label{fig:iris_uae}
\end{figure}

The success of automated border control systems used in the United Arab Emirates~\cite{AlRaisi-UAEIris-Elsevier-2008}, where it is common for individuals to conceal a substantial part of their face on account of religious beliefs, serve to demonstrate the robustness of iris recognition systems to face coverings. In these scenarios, such as that illustrated in Fig.~\ref{fig:iris_uae}, whereas face recognition systems generally fail completely, iris recognition systems may still perform reliably so long as the iris remains visible. They are also among the least intrusive of all approaches to biometric recognition. This would suggest that, at least compared to face recognition counterparts, the reliability of iris recognition systems should be relatively unaffected as a consequence of mask wearing in the COVID-19 era. 

\footnotetext[5]{Source: \url{https://en.wikipedia.org/wiki/File:IrisGuard-UAE.JPG}}

It is worth mentioning that the usefulness of the anatomy of the human eye with regard to biometrics is not limited to the irides. For example, the retinal blood vessels are suitable for the purposes of biometric recognition. However, retinal imaging requires close proximity of a highly cooperative data subject to the specialised acquisition device which sends a beam of light inside the eye to fully illuminate the retina (see e.g. \cite{Lajevardi-Retina-2013}). Although retinal structures exhibit a high degree of distinctiveness and hence good biometric performance, the need for a specialised sensor and the perceived intrusiveness of the acquisition process have been considered as obstacles to adoption of this biometric characteristic. The blood vessels present in the ocular surface have also been shown to exhibit some discriminative power and hence suitability of biometric recognition \cite{Rot-ScleraVein-2020}. The acquisition process for those, albeit less arduous than for the retinal images, still requires a high-resolution camera and subject cooperation in gazing in the required directions. Thus far, however, biometric recognition with ocular vasculature received relatively little attention beyond academic studies.

\subsection{Periocular recognition and soft-biometrics}
\label{subsec:influence_periocular}


Periocular recognition, namely recognition observing biometric characteristics from the area surrounding the eye~\cite{Alonso-PRL-PeriocularSurvey-2016}, offers potential for a compromise between the respective strengths and weaknesses of face and iris recognition systems. Unlike face recognition, periocular recognition can be reliable even when substantial portions of the face are occluded (opaque masks) or distorted (transparent masks). Unlike iris recognition, periocular recognition can be reliable in relatively unconstrained acquisition scenarios. Compared to alternative ocular biometrics, periocular recognition systems are also less demanding in terms of subject cooperation.
 

Due to those and other properties, periocular recognition was explored extensively during the last decade. Similarly to work in iris recognition, much of it has direct relevance to biometrics in the COVID-19 era, in particular with regards the wearing of face masks. In fact, one of the most popular use cases thus far for periocular recognition involves consumer mobile devices~\cite{Raja-BIOSIG-SmartphonePeriocular-2014,deFreitas-BTAS-PeriocularMobile-2015} which can readily capture high quality images of the periocular region with onboard cameras. This approach to biometric recognition, e.g.\ to unlock a personal device, is of obvious appeal in the COVID-19 era when masks must be worn in public spaces and where tactile interactions, e.g.\ to enter a password or code, must preferably be avoided. 

In most works, reliable verification rates can be achieved by extracting features from the periocular region. However, the error rates are not yet as good as those yielded by face verification schemes under controlled scenarios. Nevertheless, the periocular features can be used to improve the performance of unconstrained facial images as shown in \cite{deFreitas-BTAS-PeriocularMobile-2015}. Similarly, Park~\textit{et al.}~showed in~\cite{Park-TIFS-periocularVIS-2010} how the rank-1 accuracy was multiplied by a factor of two in a similar scenario using a synthetic dataset of face images treated artificially to occlude all but the face region above the nose. In other words, the success chances of correctly identifying a person within a group are doubled when the periocular information is analysed in parallel to the global face image. Some newer works have also explored the fair of these methods across gender \cite{Krishnan-GenderFairnessPeriocular-ICPR-2021}, reporting an equivalent performance of males and females for ocular-based mobile user-authentication at lower false match rates.

In addition to the aforementioned works, some multimodal approaches combining face, iris, and the periocular region have been proposed for mobile devices \cite{Raja-ICB-SmartphonePeriocularFaceIrs-2015}, also incorporating template protection in order to comply with the newest data privacy regulations such as the European GDPR \cite{Stokkenes-IPTA-FacePeriocularBFSmartphone-2016}.

As pointed out in Sect.~\ref{subsec:influence_iris}, in such uncontrolled conditions where the iris cannot always be used due to a low quality or resolution of the samples, that lack of quality of acquired biometric information can be addressed using super-resolution. Even though some approaches have already been proposed for the periocular region, based mostly on deep learning models \cite{Ipe-ICACN-PeriocularSRCNN-2019,Tapia-WIFS-SRIris-2020}, there is still a long way ahead before they are deployed in practical applications.

In addition to providing identity information, facial images can also be used to extract other soft biometric information, such as age range, gender, or ethnicity. Alonso-Fernandez \textit{et al.}~benchmarked the performance of six different CNNs for soft-biometrics. Also for this prupose, the results obtained indicate the possibility of performing soft-biometrics classification using images containing only the ocular or mouth regions, without a significant drop in performance in comparison to using the entire face. Furthermore, it can be observed in their study how different CNN models perform better for different population groups in terms of age or ethnicity. Therefore, the authors indicated that the fusion of information stemming form different architectures may improve the performance of the periocular region, making it eventually similar to that of unoccluded facial images. Similarly, the periocular region can be also utilised to estimate emotions using handcrafted textural features \cite{Alonso-Fernandez-SITIS-PeriocularExpression-2018} or deep learning 
\cite{Reddy-IJCNN-DeepPeriocularExpression-2020}.

\subsection{Voice recognition}


Progress in voice recognition has been rapid in recent years~\cite{kinnunen2010overview,hansen2015,Todisco-ASVIS-2016,nagrani2017voxceleb,snyder2018x}. 
Being among the most convenient of all biometrics technologies, voice recognition is now also among the most ubiquitous, being used for verification across a broad range of different services and devices, e.g.\ telephone banking services and devices such as smart phones, speakers, and watches that either contain or provide access to personal or sensitive data.

The consequences of COVID-19 upon voice recognition systems depend largely on the effect of face masks on the production of speech. Face masks obstruct the lower parts of the face and present an obstacle to the usual transmission of speech sounds; they interfere with the air pressure variations emanating from the mouth and nose. The effect is similar to acoustic filters such as sound absorbing fabrics used for soundproofing or automobile exhaust mufflers~\cite{automobile}. Since masks are designed to hinder the propagation of viral particles of sub-micron size, typically they consist of particularly dense fabric layers. The effect on speech is an often-substantial attenuation and damping. A study on the impact of fabrics on sound is reported in~\cite{seddeq2013investigation,TANG2017360}, which shows how acoustic effects are influenced by the particular textile and its thickness, density and porosity. Denser structures tend to absorb sound at frequencies above 2~{kHz}, while thicker structures absorb sound of frequencies below 500~{Hz}. With these bands overlapping that of human speech, masks attenuate and distort speech signals and hence degrade the reliability of voice biometric systems that are trained with normal (unmasked) speech. 

Masks can also have a negative impact on presentation attack detection (PAD) systems, which present countermeasures to discriminate bonafide vs spoofed speech. These systems are based on spectral features obtained from the two classes. It becomes clear that any modification/deviation of the bonafide spectrum results in greater difficulty in detecting it. Moreover, other countermeasure systems are based on the detection of the POP noise~\cite{pop}: a bonafide user emits pop noise which naturally incurred while speaking close to the microphone. This noise is attenuated by the mask and, consequently, PAD performance decreases.

\begin{figure}[t]
\centering
\subfloat[mask-free]{\includegraphics[width=0.7\linewidth]{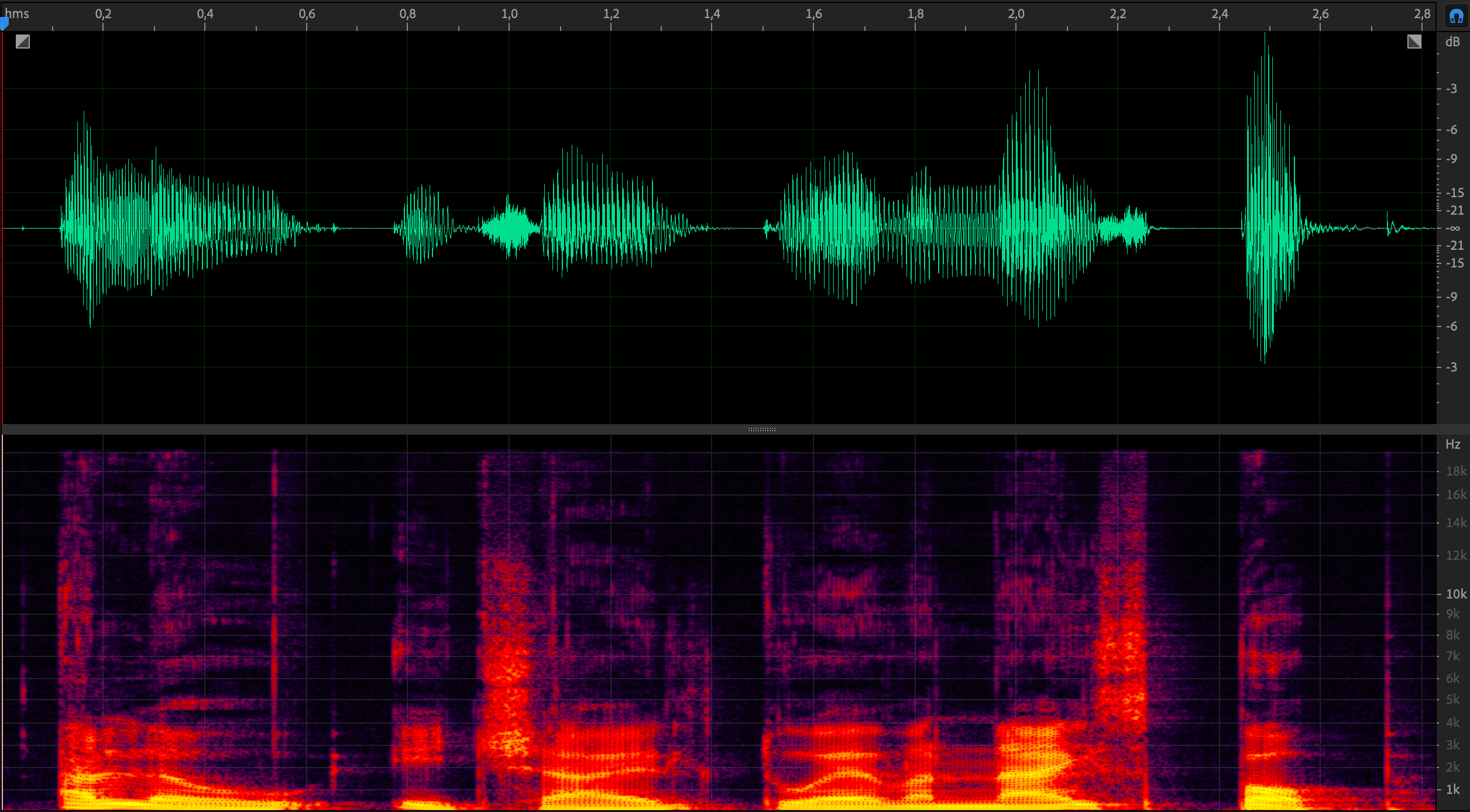}} \hfill 
\subfloat[surgical mask]{\includegraphics[width=0.7\linewidth]{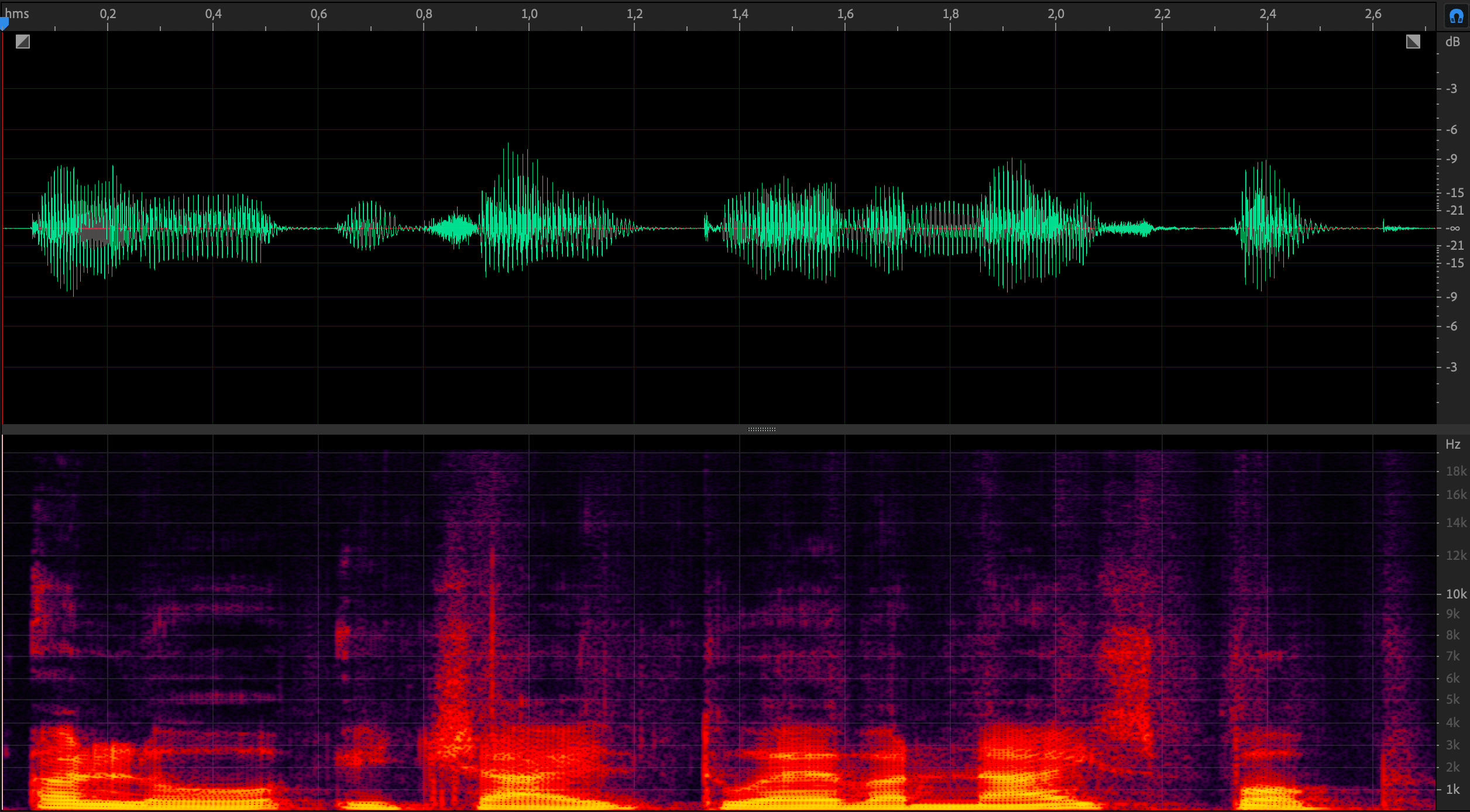}} \hfill \\
\subfloat[cloth mask]{\includegraphics[width=0.7\linewidth]{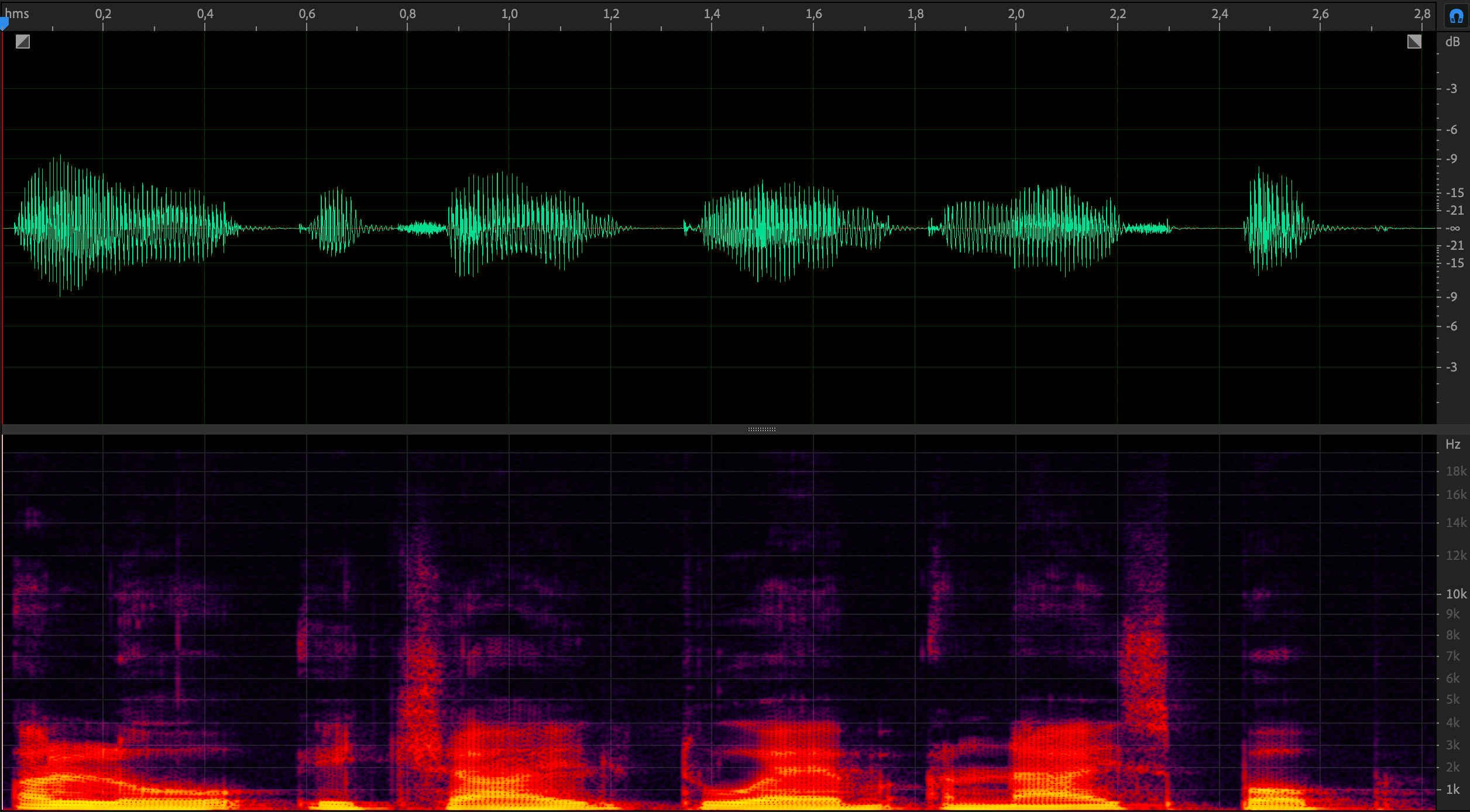}} \hfill 
\subfloat[dense cloth mask]{\includegraphics[width=0.7\linewidth]{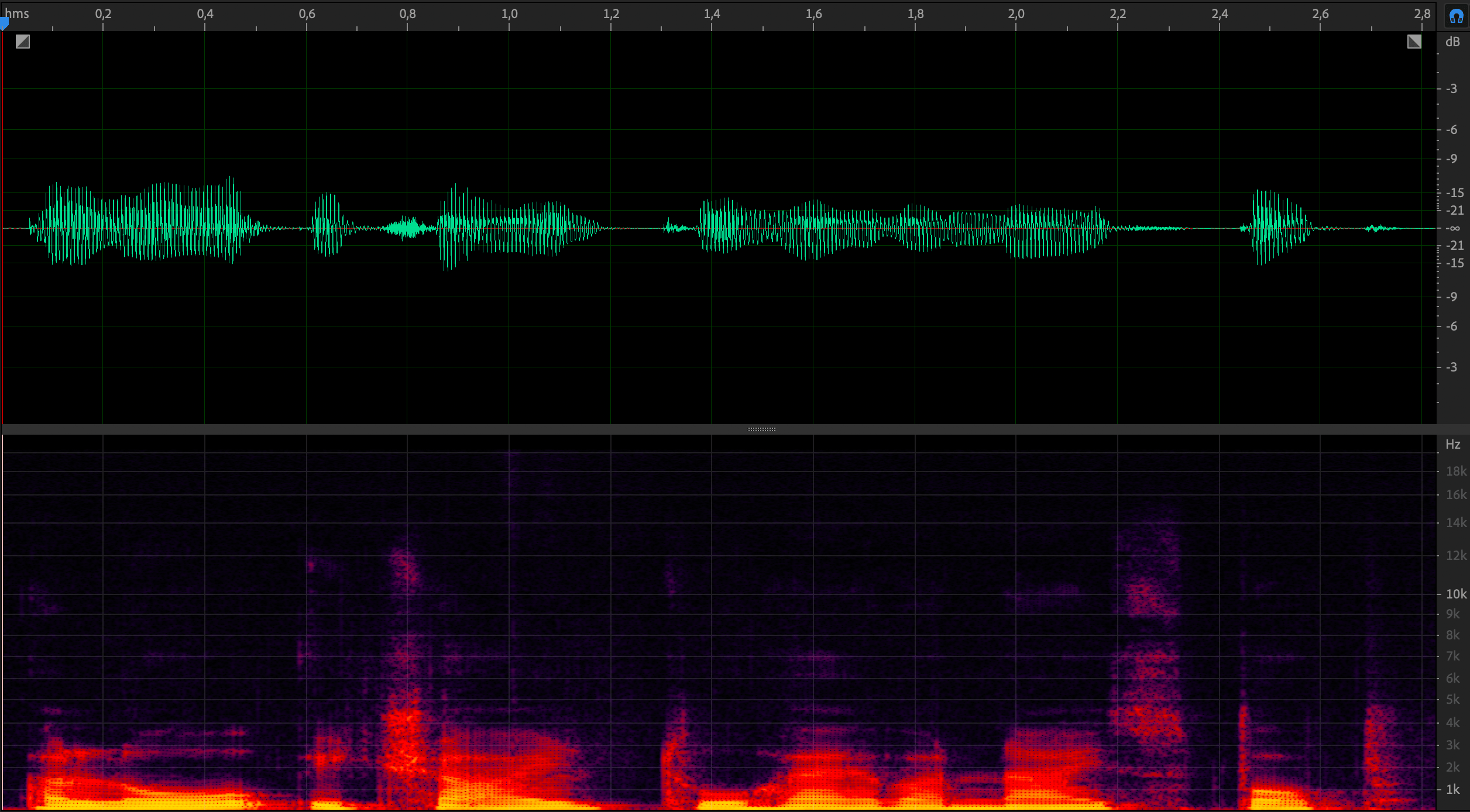}}
\caption{Examples of four spectrograms of the utterance: \textit{allow each child to have an ice pop}, pronounced by the same speaker wearing different types of masks: (a) mask-free, (b) surgical, (c) cloth and (d) dense cloth mask.}\vspace*{-0.5cm}
\label{fig:speech_fig}
\end{figure}

Fig.~\ref{fig:speech_fig} shows speech waveforms and corresponding spectrograms derived using the short-time Fourier transform (STFT) for four different recordings of read speech. The text content is identical for all four recordings: \textit{allow each child to have an ice pop}. The first is for a regular, mask-free recording while the other three are for the same speaker wearing a surgical mask, a thin or light cloth mask and a dense cloth mask. Note that the word \textit{pop} pronounced at the end of the sentence becomes less and less noticeable as you wear heavier masks. Another notable effect concerns the attenuation of high frequencies for heavier masks, which affects not only recognition performance but also speech intelligibility~\cite{Mac2019}. 

Related to these aforementioned issues, a study of the impact of face coverings upon the voice biometrics is reported in~\cite{Saeidi+2016}. It assessed and analysed the acoustic properties of four coverings (motorcycle helmet, rubber mask, surgical mask and scarf). The impact of all four coverings was found to be negligible for frequencies less than 1~kHz, while substantial levels of attenuation were observed for frequencies above 4~kHz; 4~kHz is not a general mark, since peaks at 1.8 kHz are reported for some masks. 
Face coverings were shown to degrade the accuracy of an i-vector/PLDA speaker recognition system. However, the treatment of speech data with inverted mask transfer functions was shown to improve accuracy to a level closer to the original. 
Similarly, face masks distort speech data above 4 kHz. The degradation to performance, however, is modest since the substantial effects are at higher frequencies where speech energy (and discriminative biometric information) is typically lower than it is at lower frequencies where the effects are much milder.


To reflect the current issues in the voice biometrics community, the 2020 findings of the $12^{\mbox{th}}$ Computational Paralinguistics Challenge (COMPARE) considered a mask detection sub-challenge. 
System fusion results for the challenge baselines show that the task is far from being solved. Speech signals, in this context, are not only relevant to voice biometrics but are usable to detect signal distortions.

The existing work stands to show that facial masks do affect voice-based technologies, and there is potential to compensate these effects. Thus the relevance of speaker recognition increases in this time, since it is unintrusive and touchless , that is, it can be done at distance, without any physical interaction (over the phone).

\section{Remote and mobile biometric recognition}
\label{sec:remote_biometric_authentication}
The COVID-19 pandemic has caused disruptions to many aspects of life. As a result of physical interactions being necessarily limited or even forbidden, many have had no alternative but to work remotely or to receive education online. With authentication being needed to access many services and resources, and without the possibility of physical means to identification, the deployment of biometric solutions for remote authentication has soared in recent times~\cite{Burt-COVID-2020}.
Remote biometric authentication has already attracted significant attention~\cite{HandbookRemote,Guo-Mobile-2017} and is already being exploited for, e.g.
eBanking, eLearning, and eBoarders. With an increasing percentage of personal mobile devices now incorporating fingerprint, microphone and imaging sensors, remote biometric authentication is deployable even without the need for costly, specialist or shared equipment. The latter is of obvious appeal in a pandemic, where the use of touchless, personal biometric sensors and devices can help reduce spread of the virus.

Some specific biometric characteristics lend themselves more naturally to remote authentication than others. They are dictated by the level of required user cooperation and the need for specialist sensors.
Face, voice, and keystroke/mouse dynamics are among the most popular 
characteristics for remote biometric authentication~\cite{Kaur16,FENU201883}. These characteristics can be captured with sensors which are likely to be embedded in the subjects' devices, e.g. camera, microphone, keyboard and mouse. 
As discussed in the following, remote biometric authentication entails a number of specific challenges related to 
mobile biometrics, remote education, as well as security and privacy.

\subsection{Mobile biometrics}
\label{sec:mobile}

The ever-increasing number of smartphones in use today has fueled research in mobile biometric recognition solutions, e.g.\ mobile face recognition~\cite{RATTANI-mobileface-2018} and mobile voice recognition~\cite{Khoury-Mobile-Speaker-2013,gomar2015system,bisio2018smart}. Numerous biometric algorithms specifically designed or adapted to the mobile environment have been proposed in the literature~\cite{Rattani-Selfie-2019}. Additionally, commercial solutions for mobile biometric recognition based on inbuilt smartphone sensors or hardware/software co-design are already available.


Proposed solutions can be categorized depending on where the comparison of biometric data takes place:

\begin{itemize} 
\item Biometric comparison is performed on the client side,
as proposed by the 
Fast IDentity Online (FIDO) Alliance~\cite{fido}. 
An advantage of this scheme is that biometric data is kept on the user device, leading to improved privacy protection. On the other hand, 
users may require specific sensors and 
installed software to enable 
authentication.
\item Biometric comparison is performed on the server side. These comparisons depend upon the secure transmission of biometric data (see Section~\ref{sec:remote_secpriv}), with relatively little 
specific software being required on the user device. 
\end{itemize}

One limiting factor of mobile biometrics stems from processing complexity and memory footprints. Whereas server side computation capacity and memory resources are typically abundant, mobile devices resources running on battery power are relatively limited. Many state-of-the-art biometric recognition algorithms are based on large (deep) neural networks which require a large amount of data storage and are computationally expensive, thereby prohibiting their deployment on mobile devices. 
This has spurned research in efficient, and low footprint approaches to biometric computation, e.g.\ using smaller, more shallow neural networks~\cite{Ba-2014}. A number of different approaches to compress neural networks have been proposed, e.g.\ based on student-teacher networks~\cite{Luo-2016} or pruning~\cite{Molchanov-pruning-2017}. These approaches trade model size and inference time against system performance. However, this trade-off still has to be optimized for mobile systems, while the implications of limited resources extend to other biometric sub-processes too, e.g.\ PAD. 

In summary, mobile biometric authentication clearly has a role to play in the COVID-19 era. 
Touchless, personal mobile biometrics solutions can help to deliver reliable authentication while also meeting strict hygiene requirements, even if the efficient integration of biometric recognition technologies into mobile device platforms remains challenging.

\subsection{Biometrics in remote education}\label{sec:elearning}

The use of learning management systems has increased dramatically in recent years, not least due to the promotion of home-schooling and eLearning during the COVID-19 pandemic. Learning management systems deliver remote education via electronic media. eLearning systems often require some form of identity management for the authentication of remote students. Biometrics solutions have proved extremely popular, with 
a number of strategies to integrate biometric recognition in eLearning environments having been proposed in recent years~\cite{eLearniingsurvey,Sanna-2017}.

\begin{figure}[t]
\centering
\includegraphics[width=0.75\linewidth]{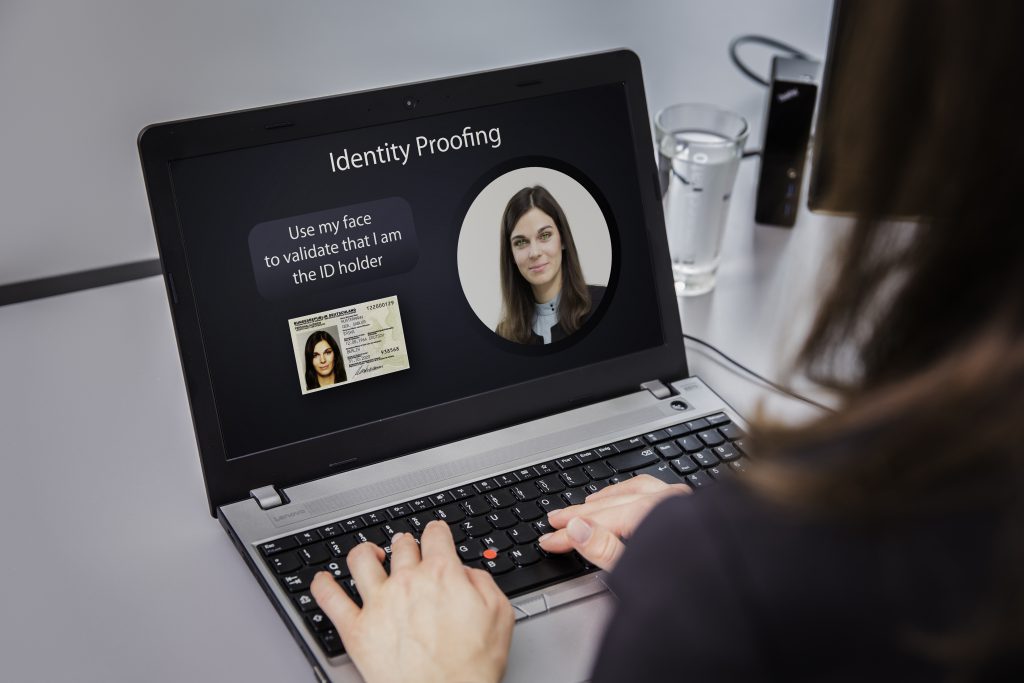}
\caption{BioID\textsuperscript{\tiny\textregistered}
 Identity Proofing for e-learning platforms \cite{bioid-2020}}\vspace*{-0.5cm}
\label{fig:eLearning}
\end{figure}

In the eLearning arena, biometric technologies are used for user login, user monitoring, attention or emotion estimation, and authorship verification. Fig.~\ref{fig:eLearning} shows an example for user login to an eLearning platform.
Both one-time authentication (biometric verification at a single point in time) and continuous authentication (periodic over time) have utility in eLearning scenarios. Whereas one-time authentication might 
be suitable to authenticate students submitting homework, 
continuous authentication may be preferred to prevent students cheating while sitting remote examinations~\cite{Flior10}. 
In order to minimise inconvenience, continuous biometric authentication calls for
the use of biometric characteristics which require little to no user cooperation~\cite{eLearniingsurvey}, e.g.\ text-independent keystroke dynamics~\cite{Morales_Keystroke,Bours17}. 

Presentation attacks can present a substantial threat to biometric technologies deployed in such scenarios (see Section~\ref{sec:remote_secpriv}). 
This might be why, despite significant research interest, 
only few biometric recognition systems have been deployed in operational eLearning scenarios~\cite{eLearniingsurvey}. 
Even so, 
eLearning systems will likely become more popular while the pandemic continues and, once operational, their use will likely be maintained
in the future.

\subsection{Security and privacy in remote biometrics}\label{sec:remote_secpriv}

The remote collection of biometric information gives rise to obvious security and privacy concerns; the trustworthiness of the collection environment cannot be guaranteed.
One of the potentially gravest threats in this case, especially given the absence of any human supervision (e.g.\ in contrast to the automatic boarder control use case), is that of presentation attacks or `spoofing'~\cite{Ratha-EnhancingSecurityAndPrivacy-IBM2001,Marcel-HandbookPAD-ACVPR-2019,Raghavendra-FacePAD-Survey-2017}.
Presentation attacks involve the presentation of false, manipulated or synthesized samples to a biometric system made by an attacker to masquerade as another individual.
Diverse presentation attack instruments, ranging from face masks to gummy fingers, have 
all been proved a threat.
The detection of presentation attacks in a remote setting 
can be more challenging that in a local setting, depending on whether detection countermeasures are implemented on the client side or the server side. In case PAD is performed on the client side, hardware-based detection approaches can be employed, though these require specific, additional equipment beyond those used purely for recognition. Even these approach might still be vulnerable to presentation attacks, as demonstrated for Apple's Face ID system~\cite{iPhone-2017}. If PAD is implemented on the server side, then software-based attack detection mechanisms represent the only solution. Such software-based PAD for remote face and voice recognition were explored in the EU-H2020 TeSLA project~\cite{Bhattacharjee18}. It is expected that more research will be devoted to this topic in the future \cite{DBLP:conf/fgr/FangBKD21,DBLP:journals/pr/FangDKK22}.

In addition to the threat of direct attacks performed at the sensor level, there is also the possibility of indirect attacks performed at the system level.
The storage of personal biometric information on mobile devices as well as the transmission of this information from the client to a cloud based server calls for strong data protection mechanisms. 
While traditional encryption and cryptographic protocols can obviously be applied to the protection of biometric data, any processing applied to the data required prior decryption, which still 
leaves biometric information vulnerable to interception.
Encryption mechanisms designed specifically for biometric recognition in the form of template protection~\cite{Rathgeb-BTP-Survey-EURASIP-2011} overcome this vulnerability by enabling comparison of biometric data in the encrypted domain. Specific communication architectures that ensure privacy protection in remote biometric authentication scenarios where biometric data is transmitted between a client and a server have already been introduced, e.g.\ the Biometric Open Protocol Standard (BOPS)~\cite{BOPS} which supports the homomorphic encryption~\cite{Moore-HE-Survey-ISCAS-2014} of biometric data. 

As it has been described in this section, the use of remote biometric authentication in the times of COVID-19 provides many advantages. However, in order to achieve trustworthy identity management, it also requires appropriate mechanisms to protect privacy. Countermeasures to prevent or detect presentation attacks are also essential. The latter is usually more challenging in a remote authentication scenario, where means of detecting attacks may be more limited compared to conventional (accessible) biometric systems.

\section{Emerging technologies}
\label{sec:emerging_technologies}
As discussed in the previous sections, the COVID-19 pandemic poses specific challenges to biometric technologies.
However, it is also expected to foster research and development in emerging biometrics characteristics which stand to meet new requirements relating to the pandemic, as well as the use of biometric information directly for virus detection and monitoring 
e.g.\ of infected individuals. Such emerging biometric technologies are described in the following. 




\subsection{Touchless, hand-based biometrics}
\begin{figure}[t]
\centering
\subfloat[dry]{\includegraphics[height=3.75cm]{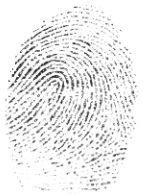}} \hfil
\subfloat[normally moist]{\includegraphics[height=4cm]{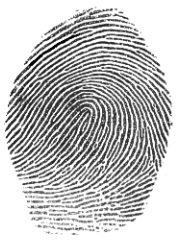}}
\caption{Example of a dry fingerprint and the same fingerprint with normal moist (taken from \cite{Olson-Moisture-2015}).}
\label{fig:dryfp}
\end{figure}

Hydro-alcoholic gel, strongly advocated as a convenient means to disinfection during the COVID era, can be used to protect the users of touch-based sensors such as those used for fingerprint recognition~\cite{Okereafor-JMIRBE-FingerSensorCOVID-2020}. 
While they serve to reduce sensor contamination and pathogen transmission, hydro-alcoholic gels tend to dry the skin.
The sensitivity of fingerprint sensors to variability in skin hydration is well known. It can degrade the quality of acquired fingerprints and hence also recognition reliability~\cite{Olson-Moisture-2015}.
Severe dryness can even prevent successful acquisition as illustrated in Fig.~\ref{fig:dryfp}, thereby resulting in failures to acquire.

\begin{figure}[t]
\centering
\subfloat[Stationary touchless\protect\footnotemark]{\includegraphics[width=0.45\linewidth]{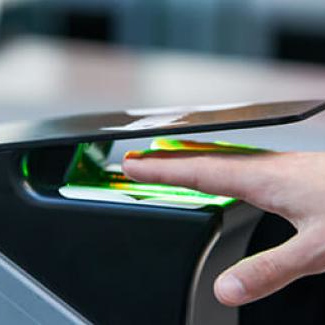}} \hfil
\subfloat[Mobile touchless]{\includegraphics[width=0.45\linewidth]{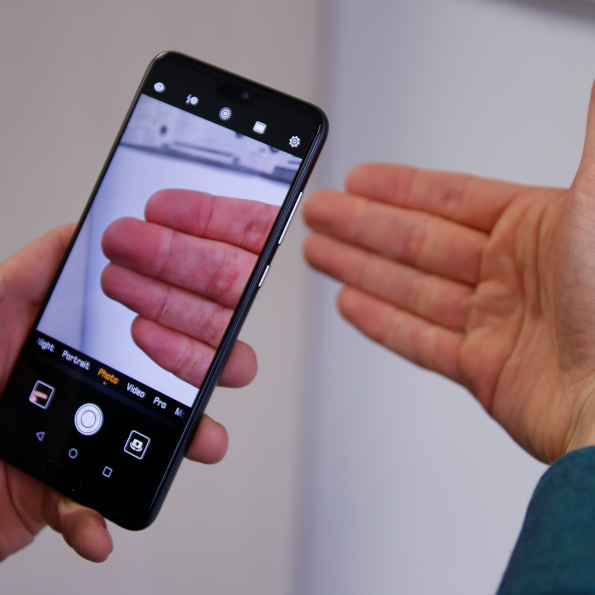}}
\caption{Touchless capturing of fingerprints}\vspace*{-0.5cm}
\label{fig:touchless}
\end{figure}
\footnotetext[7]{Source: \url{https://pbs.twimg.com/media/DyCFi_AWsAMN8MK.jpg}}


Hygiene concerns 
have increased societal resistance to the use of touch-based sensors. 
These concerns have in turn fueled research efforts in 2D or 3D touchless fingerprint recognition systems~\cite{Priesnitz-2dfinger-Survey-2020,Ajay-3Dfinger-2018} such as those illustrated in 
Fig.~\ref{fig:touchless}. 
Touchless fingerprint sensors are generally either
prototype hardware designs~\cite{raghavendra2014low-cost,Galbally20173D-touchless} or are adapted from general purpose devices 
adapted to 
touchless fingerprint recognition~\cite{kumar2011contactless,stein2012fingerphoto}. 

Both the capture and processing of fingerprints must usually be adapted to touchless acquisition~\cite{Priesnitz-2dfinger-Survey-2020}. 
The majority of touchless finger image acquisition sensors deliver colour images for which general image processing techniques are employed to improve contrast and sharpness. Traditional minutiae extractors and comparators may then be employed.

The interoperability of both touch-based and touchless devices is naturally desirable, e.g.\ to avoid the need for enrolment in two different systems. 
Interoperability has proven to be 
non-trivial~\cite{Libert-NIST-ContactlessFingerComp-2020,Orandi-NIST-ContactlessFingerMatcher-2020}. While some differences between the two systems, e.g. mirroring, colour-to-grayscale conversion or inverted back- and foreground, can be readily compensated for without degrading accuracy, others, e.g.\ the aspect ratio or deformation estimation, prove more challenging~\cite{Salum17touchlesstouchbased,Lin18touchlesstouchbased} and can degrade reliability. Note that fingerprint images acquired using touchless sensors do not exhibit the deformations caused by pressing the finger onto a surface that characterise images acquired from touch-based sensors. Moreover, DPI alignment and ridge frequency estimation is required to enable a meaningful comparison of fingerprints acquired from touch-based and touchless sensors.


As an alternative to fingerprint recognition, some ATMs already incorporate fingervein-based recognition sensors which are robust to variability in skin hydration as well as presentation attacks. 
Images of the finger or hand are captured with NIR illumination, since light at NIR frequencies is absorbed differently by hemoglobin and the skin,
thereby allowing for the detection of vein patterns. 
Touchless fingervein and palmvein sensors have been developed~\cite{Kauba-MDPISensors-ContactlessFingerprintVein-2019,Ma-IPT-ContactlessFingervein-2019,Marattukalam-ANZCC-PalmVeinContactless-2019}, though the lack of any control in the collection process typically causes
significant rotation and translation variation.
The quality of the capturing device as well as strategies to compensate for nuisance variation are hence key to the collection of high quality images and reliable performance. Touchless capturing device designs have been presented by various researchers, e.g. in \cite{Kauba-MDPISensors-ContactlessFingerprintVein-2019}. 
This work showed that the degradation in recognition performance resulting from touchless acquisition
can be addressed using finger misplacement corrections. On the other hand, the approach presented in~\cite{Ma-IPT-ContactlessFingervein-2019} extracts a region of interest from captured samples and uses an oriented element feature extraction scheme to improve robustness.

The use of finger vein recognition for mobile devices is also emerging. 
Debiasi \textit{et al.}~developed an auxiliary NIR illumination device for smartphones
which supports the capture of hand vascular patterns~\cite{Debiasi-BTAS-NIRVeinMobile-2018}. The device is connected and controlled via
Bluetooth and can be adapted to different smartphones. The authors also presented a challenge response protocol in order to prevent replay and presentation attacks and showed that acceptable verification performance can be achieved using standard finger vein recognition algorithms.
The VeinSeek Pro app\footnote{\url{https://www.veinseek.com/}} is able to capture vein images from the hand without the need for extra hardware. This approach is based on the fact that different colors of light penetrate different depths within the skin. By removing the signal from superficial layers of the skin, the authors argue that they can more easily see deeper structures. However, to the best of our knowledge there is no analysis so far of the feasibility of using these images for vein-based biometric recognition.

In summary, in the era of the COVID-19 pandemic, touchless hand-based biometric recognition seems to be a viable alternative to conventional touch-based systems. 
These technologies achieve similar levels of performance as touch-based technologies~\cite{Priesnitz-2dfinger-Survey-2020,Ajay-3Dfinger-2018,Kauba-MDPISensors-ContactlessFingerprintVein-2019}. Some commercial products based on prototypical hardware design and general purpose devices, e.g.\ smartphones, are already available on the market. Nonetheless, touchless recognition remains an active field of research where several challenges need to be tackled, in particular recognition in challenging environmental conditions, e.g.\ uncontrolled background or varying illumination~\cite{malhotra_fingerphoto_2017,Priesnitz-2dfinger-Survey-2020}. 

\subsection{COVID detection with biometric-related technologies}




COVID-19 attacks the human body at many levels, but the damage to the respiratory system is what often proves fatal.
The production of human speech starts with air in the lungs being forced through the vocal tract. 
Diminished lung capacity or disease hence impacts upon speech production and there have been attempts to characterise the effects of COVID-19 upon speech as means to detect and diagnose infection~\cite{schuller2020covid,deshpande2020audio,bartl2020voice}. 

Initial efforts involved the collection and annotation of databases of speech as well as non-speech sounds recorded from healthy speakers and those infected with the COVID-19 virus~\cite{shuja2020covid}. The data typically includes recordings of coughs~\cite{imran2020ai4covid,brown2020exploring,sharma2020coswara}, breathing sounds~\cite{faezipour2020smartphone,trivedy2020design} as well as speech excerpts~\cite{Han2020}.

The database described in~\cite{Han2020} contains recordings of five spoken sentences and in-the-wild speech, all recorded using the Wechat App from 52 COVID-confirmed and hospitalised patients in Wuhan, China, who also rated their sleep quality, fatigue, and anxiety (low, mid, and high). After data pre-processing, 260 audio samples were obtained. 
While these early works highlight the potential of biometrics and related technology to help in the fight against the COVID-19 pandemic, 
they also highlight the need for homogenised and balanced databases which can then be used to identify more reliable and consistent biomarkers indicative of COVID-19 infection. Outcomes of these studies are very encouraging: the detection of COVID-19 through voice, but also through coughing or the sound of breathing, has an accuracy comparable to that of the antigen or saliva test~\cite{kamble21_interspeech, kamble22, das21_interspeech, avila21_interspeech}.



Thermal face imaging has also come to play a major role during the pandemic, especially for the rapid surveillance of potential infections among groups of travellers on the move, e.g.\ in airports~\cite{TFM_Airports} and shopping centres~\cite{TFM_Malls}. Thermal face images can be used to detect individuals with fever~\cite{DBLP:conf/iccvw/LinLL19}, a possible symptom of COVID-19 infection. 
Similar face captures can also be used as an alternative capture spectrum for face recognition \cite{DBLP:conf/icb/MallatDBKD19,DBLP:conf/icb/IranmaneshN19,DBLP:journals/corr/abs-1910-09524}, however, with verification performances inferior to the visible \cite{DBLP:conf/cvpr/DengGXZ19,DBLP:journals/csr/FarokhiFS16}.
Despite the ease with which thermal monitoring can be deployed, it is argued in~\cite{ThermalScanningEffec} that body temperature monitoring will be insufficient on its own to prevent 
the spread of COVID-19 into previously uninfected countries or regions and the seeding of local transmission. The European Union Aviation Safety Agency (EASA) concludes that thermal screening equipment, including thermal scanners will miss between 1\% and 20\% of passengers carrying a fever~\cite{EASAreport}. 

\section{Societal Impact}
\label{sec:societal}

As any other technology used by a large population, biometric recognition systems affect the society. So far, the positive aspects of such systems (e.g., faster authentication for border crossing or convenience for smartphone unlocking) have outweighed their disadvantages, mostly related to privacy and security issues \cite{Kindt-PrivacyIssues-Springer-2016,Tanwar-EthicsBiometrics-Springer-2019}. Such issues have been thoroughly analysed and (partially) dealt with, thereby increasing the acceptance of the users and boosting the deployment of biometric systems. Nevertheless, in the last years new, concerns have arisen related to the fairness of biometric recognition algorithms \cite{FreitasPereira-Fairness-arXiv-2021} and their trustworthiness \cite{Jain-TrustVerify-arxiv-2021,Rathgeb-DemographicFairness-arxiv-2021}. In addition, societal and ethical aspects of presentation attack detection methods have also been analysed \cite{Rebera-SocioEthicPAD-SEE-2014}.

In the context of the COVID-19 pandemic, the use of contact-based biometric systems have similarly lead to health-related concerns. Systems where contact with the capture device is necessary could still be employed in a private scenario (e.g., for unlocking your own smartphone or for remote for authentication from your own laptop), but contact-less approaches will be preferred for global applications (e.g., building access control) in order to prevent the spread of viruses. In fact, it can be argued that the use of contact-less biometrics can even reduce the transmission of pathogens in some scenarios such as airport \cite{BI-Covid19biometrics-2020}. This trend will probably remain even after the COVID-19 pandemic can be considered to be over.

On the other hand, the need for further digitalisation in almost all societal levels, including sensitive applications such as online exams or eHealth systems, where subject identification is of the utmost importance, has increased the acceptance of biometric technologies as a convenient and reliable means of authentication. Thus, more research is being done in this area \cite{Faundez-EHealthSignature-CC-2020,Vizitiu-IoTeHealthBiometrics-2021}, together with socioeconomic analysis of success and failure of big-scale implementations of such systems \cite{Effah-GhanaBiometricsSocioeconomic-ISM-2020}. 

However, further digitalisation also brings  some disadvantages. In general, and not only regarding biometric recognition, the tracking activities and health checks implemented worldwide in order to prevent the spread of COVID-19 have had deep implications on the privacy and freedom of the subjects. For instance, free travel within Schengen has been suspended for months, needing to fulfill certain criteria in terms of negative COVID-19 tests, vaccination status, or registration forms to enter a country\footnote{\url{https://reopen.europa.eu/en/}}. In addition, facial recognition systems have been used in countries such as Poland, China, or Russia to ensure that individuals in quarantine remain at home. In spite of the benefits for the collective health, \lq\lq \textit{the use of biometrics (including facial recognition) in response to COVID-19 raises a number of privacy and security concerns, particularly when these technologies are being used in the absence of specific guidance or fully informed and explicit consent. Individuals may also have problems exercising a wide range of fundamental rights, including the right of access to their personal data, the right to erasure, and the right to be informed as to the purposes of processing and who that data is shared with}'', as the Organisation for Economic Co-operation and Development (OECD) states in its policy response to Coronavirus (COVID-19) \cite{OECD-Covid19Privacy-2020}. Thus, the OECD gives a number of recommendations including the use of privacy-by-design approaches, such as the ones described in Sect.\ref{sec:remote_secpriv}, and the limitation on the time sensitive data can be stored.

The added societal concerns due to the exploitation of sensitive biometric data have been also addressed by The British Academy \cite{BA-Covid19LongSocietal-2021}. As the Academy points out, \lq\lq \textit{Sharing data is crucial for furthering research and maximising its potential to help overcome the current pandemic and better prepare for future health crises}''. However, bias or errors derived from the use of biometric technologies for authentication can result in negative impacts such as discrimination, and diminish the trust on COVID-19 related technologies. Therefore, the Academy recommends maintaining a human element in the loop. In addition, existing digital inequalities might also limit the potential benefits of health technologies and increase the social disadvantages of some groups. The report also includes some numbers: \lq\lq \textit{6 million people in the UK cannot turn on a device and up to 50\% of those are
aged under 65}''. Furthermore, in order to minimise the potential discrimination caused by biometric technologies, several characteristics should be considered: apps which rely
on voice recognition software that may not work effectively for those with a speech
impairment, can be beneficial for those with reduced sight.

In March 2022, the European Data Protection Supervisor (EDPS) published a report on COVID-19 related processing of the Union institutions, bodies, offices and agencies (EUIs) \cite{EDPS-Covid19ProcessingEUIs-2022}. In this survey, the EDPS reviews body temperature checks, contact tracing, COVID testing and handling of results, monitoring presence within the premises, vaccination campaigns, access control, and the use of IT-tools in telework. Regarding access control, where biometric recognition systems can be in place, the EUIs correctly informed the individuals about the processing activities carried our and specified a time limit for data retention, as recommended by the OECD. However, as the report points out, the lawful grounds of this identification requirement may not be given, since \lq\lq \textit{staff members [...] cannot provide freely given, specific, informed and unambiguous as well as explicit consent}''. Similarly, \lq\lq \textit{consent would also not be appropriate for visitors, who are in most cases obliged to come to the EUI premises for work purposes}''. Also, some EUIs had not indicated that they process health data even if they were doing so. In view of these negative impact on the privacy rights of the individuals, the EDPS recommends the EUIs to check the lawfulness and regularly reassess the necessity and proportionality of the existing COVID-related processing activities. 

From those reports we can conclude that biometrics and other technologies have not only provided the subjects with additional advantages to access digital services, but have also had a negative impact on their right to privacy. Thus, we would like to urge the community to assess the necessity of identity checks before implementing them, and use all the available tools to minimise the negative impact of such a control: biometric template protection schemes to prevent sensitive data leakage, or presentation attack detection modules to minimise the success chances of identity theft.

\section{Conclusions}
\label{sec:conclusions}

This article has summarised the main challenges posed by the pandemic to biometric recognition, as well as the new opportunities for existing biometrics to be harnessed or adapted to the COVID-19 era, or where biometrics technology itself has potential to help in the fight against the virus. The use of hygienic masks covering the nose and mouth, as well as the secondary impacts of strict hygiene measures implemented to control the spread of pathogens all have potential to impact upon biometrics technology, thereby
calling for new research to maintain reliable recognition performance.

Facial biometrics are among the most impacted characteristic; masks occlude a considerable part of the face, 
leading to degraded recognition performance.
This is the case not only for opaque masks but also for transparent face shields, since reflections caused variation that is non-trivial to model. Opportunities to overcome these difficulties are found by focusing parts of the face that remain uncovered, namely the iris and the wider periocular region. 

Whereas solutions to iris recognition that use the NIR spectrum are well studied, numerous efforts in recent years have focused on less constrained approaches to iris recognition that use mobile devices and the visible spectrum. Given the lower quality of such images, image super-resolution techniques have been proposed to improve image quality. Such techniques can also be applied to the full periocular region. To date, the adoption of such systems is low, but likely to increase in the future.

Hand-based biometric systems are also affected by the new hygiene practices which typically result in drier skin, lower quality fingerprint images and degraded recognition performance. Both touch-based and touch-less systems are affected. Vein-based recognition systems are more robust to variations in skin condition. In contrast to traditional touched-based vein sensors, touch-less capture devices introduced in the last two years can reduce the risk of infection from contact with a contaminated surface. Further research is nonetheless needed to bridge the gap between the performance of less constrained, touchless systems and their better constrained touch-based counterparts. 

Like facial biometrics, voice biometric systems are also impacted by the wearing of facial masks which can interfere with speech production. Like many other forms of illness, COVID-19 infections can also interfere with the human speech production system and also degrade recognition performance. 
These same effects upon the speech production mechanism, however, offer potential for the detection of pulmonary complications such as those associated with serious COVID-19 infections.

Still, the challenges in ensuring reliable biometric recognition performance have grown considerably during the COVID-19 era and call for renewed research efforts. With many now working or receiving education at home, some of the greatest challenges relate to the use of biometric technology in remote, unsupervised verification scenarios.
This in turn gives greater importance to continuous authentication, presentation attack detection, or biometric template protection to ensure security and privacy in such settings which have come to so define the COVID-19 era. 



\bibliographystyle{IEEEtran}
\bibliography{references}

\end{document}